\documentclass[conference]{IEEEtran}

\usepackage{graphicx}

\usepackage{slashbox}

\usepackage[cmex10]{amsmath}

\begin{document}
\title{\bf Review on Telemonitoring of Maternal Health care Targeting Medical Cyber-Physical Systems}


%\author{\IEEEauthorblockN{Md. Hanif Seddiqui}
%\IEEEauthorblockA{Dept. of Computer Science and Engineering\\
%University of Chittagong\\
%Chittagong - 4331, Bangladesh\\
%Email: hanif@cu.ac.bd}
%\and
%\IEEEauthorblockN{Abdullah Al Mohammad Maruf}
%\IEEEauthorblockA{Dept. of Computer Science and Engineering\\
%University of Chittagong\\
%Chittagong - 4331, Bangladesh\\
%Email: maruf.cu.cse09@gmail.com}
%
%
%}

\author{
	\IEEEauthorblockN{
		Mohammod Abul Kashem\IEEEauthorrefmark{1}\IEEEauthorrefmark{3},
		Md. Hanif Seddiqui\IEEEauthorrefmark{2}\IEEEauthorrefmark{3},
		Nejib Moalla\IEEEauthorrefmark{3}, 
		Aicha Sekhari\IEEEauthorrefmark{3}, 
		Yacine Ouzrout\IEEEauthorrefmark{3}
	}
\IEEEauthorblockA{\IEEEauthorrefmark{1}Dept. of Computer Science and Engineering\\
Dhaka University of Engineering and Technology, Dhaka, Bangladesh}\\
\IEEEauthorblockA{\IEEEauthorrefmark{2}Dept. of Computer Science and Engineering\\
University of Chittagong, Chittagong, Bangladesh}\\
\IEEEauthorblockA{\IEEEauthorrefmark{2}DISP Laboratory\\
IUT Lumiere Lyon 2, Lyon, France}\\
Email: 	drkashemll@duet.ac.bd,
		hanif@cu.ac.bd,
		Nejib.Moalla@univ-lyon2.fr,
		aicha.sekhari@univ-lyon2.fr,
		yacine.ouzrout@univ-lyon2.fr\\
}

\maketitle

\begin{abstract}
%objective
We aim to review available literature related to the telemonitoring of maternal health care for a comprehensive understanding of the roles of Medical Cyber-Physical-Systems (MCPS) as cutting edge technology in maternal risk factor management, and for understanding the possible research gap in the domain. 
%method
In this regard, we search literature through google scholar and PubMed databases for published studies that focus on maternal telemonitoring systems using sensors, Cyber-Physical-System (CPS) and information decision systems for addressing risk factors. 
%result
We extract 1340 articles relevant to maternal health care that addresses different risk factors as their managerial issues. Of a large number of relevant articles, we included 26 prospective studies relating to sensors or Medical Cyber-Physical-Systems (MCPS) based maternal telemonitoring. Of the 1340 primary articles, we have short-listed 26 articles (12 articles for risk factor analysis, 9 for synthesis matrices and 5 papers for finding essential elements. We have extracted 17 vital symptoms as maternal risk factors during pregnancy. Moreover, we have identified
a number of cyber-frameworks as the basis of information decision support system to cope with the different maternal complexities. 
%conclusion
We have found the Medical Cyber-Physical System (MCPS) as a promising technology to manage the vital risk factors quickly and efficiently by the care provider from a distant place to reduce the fatal risks. Despite communication issues, MCPS is a key-enabling technology to cope with the advancement of telemonitoring paradigm in the maternal health care system. 
\end{abstract}

\section{Introduction}
\label{sec:intro}
%millenium sustainable development
The fact sheets of World Health Organization (WHO)~\footnote{http://www.who.int/mediacentre/factsheets/fs348/en/} contain the real scenario of the present maternal mortality status as a partial fulfillment of the Millennium Development Goals (MDG) 4 and 5. WHO announces a new agenda, called Sustainable Development Goal (SDG), as a follow-up of MDG. The key target of the new agenda is to reduce the global maternal mortality ratio to less than 7 per 100,000 between 2016 and 2030. Now it is getting many researchers attention to contribute to the reduction of maternal mortality rate through introducing new technologies in the pathway.

With the recent advancement of Information and Communication Technology (ICT), especially wireless 
sensor networks (WSN), medical sensors and cloud computing, heterogeneous data integration and automation are getting researchers attention in the health care domain, although isolated sensors (e.g. ECG, USG, Echo etc.) are being used in this domain since long ago. Sensors usually produce lots of data instantaneously with important and unimportant data as well. Automation often requires filtering unimportant data at the physical level to ease communication technology to transfer only important data to be used in the cyber (computer) level. This technique is coined as Cyber-Physical-Systems (CPS). CPS research is revealing numerous opportunities and challenges in medicine and biomedical engineering~\cite{baheti2011cyber}. CPS in medicine and biomedical engineering is often called as Medical CPS (MCPS).

Cyber-Physical-Systems (CPS) was identified as a key research area by the US National Science Foundation (NSF)  in 2008 and was listed as the number one research priority by the US President’s Council of Advisors on Science and Technology~\cite{wang2011secured}. The advancement of WSN, medical sensors and cloud computing may enable CPS a powerful candidate for in-home patient care~\cite{milenkovic2006wireless}.

The study of CPS (often called as MCPS) in maternal health care for telemonitoring is still in its early stage of development. However, CPS has been introduced in general health care domain~\cite{haque2014review}. As health care mission is critical services, we need to focus on the safety and security issue in this domain~\cite{banerjee2012ensuring}. Moreover, maternal health care systems heavily affected by the risk factor identification and integration to produce an accurate alarm. Many issues are still open for future research such as real-time processing, efficient data query, storage management, and security. Therefore, one of the aims of this review paper is to indicate some unanswered questions or research gap in CPS for health care.
  
%Paper organization
The rest of the paper is organized as follows. 
%\textbf{Section~\ref{sec:gen-term}} focuses on the definition of general terminology which are necessary to comprehend the rest of the sections.
\textbf{Section~\ref{sec:material}} adheres the literature search, study selection, data extraction and analysis.
\textbf{Section~\ref{sec:mat-health}} contains maternal health-related detailed literature review that includes risk factors, synthesis matrix and other literature contribution whereas discussion on some interesting finding has been articulated in \textbf{Section~\ref{sec:discussion}}. 
Concluded remarks of our work are elaborated in \textbf{Section~\ref{sec:conclusion}}.
 
\section{Materials and Methods}
\label{sec:material}
During conducting this review, we adhere to be focused on our targeted keywords: maternal health care, maternal risk factors, CPS or sensor, and telemonitoring.

\subsection{Literature Search and Study Selection}

We have conducted a comprehensive search of google scholar and PubMed
databases for extracting literature on CPS/sensor based telemonitoring 
system considering maternal risk factors from 2010 through 2015. During 
our search, we have used free text (e.g. CPS sensor maternal risk factors)
to maximize the coverage. We have customized our search by putting range
from 2010 through 2015 that includes patent and citation in google scholar.
We have found 1340 articles from the search in the
google scholar~\footnote{https://scholar.google.fr}, however,
no article has been found in PubMed~\footnote{www.ncbi.nlm.nih.gov/pubmed/}.

\begin{figure}
  \includegraphics[width=\linewidth]{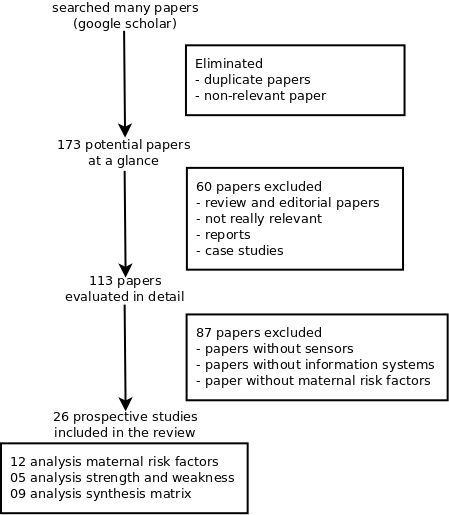}
  \caption{Flow diagram for literature search and study selection}
  \label{fig:study-selection}
\end{figure}

Fig.~\ref{fig:study-selection} summarizes our process of literature search and
study selection. Initially, we have found 1340 articles through google scholar
with the free query text: CPS sensor maternal risk factors. At the primary scrutinizing,
we have eliminated duplicate and non-relevant or non-technical articles. 
We have detected 173 relevant 
articles. We have focused deeply on 173 articles and we have again found some articles
as review or editorial articles, some as reports and case studies and a few are not 
really focused on our area of interest. We have detected 60 such type of articles.
Then we have remained 113 articles. We have gone through each of the articles to 
detect interesting findings. Furthermore, our study has depicted 87 papers non-relevant
due to the lack of CPS/sensor, information systems or maternal risk factor based study.
Finally, we have 26 prospective studies that focus on our concentrated field of interest.

\subsection{Data Extraction and Analysis}
The final 26 articles are classified into three groups: 1. 12 papers are studied to identify
risk factors associated to maternal health care, 2. 5 articles are analyzed to focus the 
strength and weakness to pinpoint the research gap, and 3. 9 articles are analyzed
to find attributes that can characterize any other literature.

We have assessed 12 articles to identify risk factors and we have found the following risk
factors evidently: age, body mass index(BMI), blood pressure, blood oxygenation, blood glucose, body temperature, physical activity, maternal ECG, nausea at the first-trimester, vaginal discharge at the first-trimester, maternal serum alpha-fetoprotein level (MSAFP) at the second-trimester, contraction at the third-trimester, abnormal fetal position, electrical uterine activity, fetal heart activity, and 
fetal movement activity.

In the second group of 5 articles, we have noticed authors' claims and considered as their strength to find some interesting key-enabling ideas.

We have detected a number of attributes to characterize any relevant articles. The attributes are 
as follows: CPS, sensor technology, telemonitoring, interoperability, S3 (safety, security and sustainability), adaptability, reliability, cyber repository or storage, external or environmental factors, communication technology, DMS, risk factors, physical activity, behavioral factors, privacy issues, maternal monitoring and fetal monitoring.

\section{Maternal Health Review Result}
\label{sec:mat-health}

\subsection{Risk Factors}
As we concentrate our study on maternal health care, we look for fundamental causes or risk factors of maternal health care. The risk factors include age, body mass index(BMI), blood pressure, blood oxygenation, blood glucose, body temperature, physical activity, maternal ECG, nausea at the first-trimester, vaginal discharge at the first-trimester, maternal serum alpha-fetoprotein level (MSAFP) at the second-trimester, contraction at the third-trimester, abnormal fetal position, electrical uterine activity, fetal heart activity, fetal movement activity and so on. All the factors have threshold values or measurement units. Medical professionals or researchers research to find out the different factors for further medication, clinical diagnosis or other management. Due to the different factors, a pregnancy can be classified as normal, moderate or high-risk pregnancy by an obstetrician~\cite{gorthi2009automated}.

In the paper~\cite{lampinen2009review}, authors study on the range of maternal ages and find out the lower and upper bound of age for high-risk pregnancy. They have shown that the advanced maternal age is associated with a range of adverse pregnancy outcomes. Moreover, a study shows a method for comparing age factor of a pregnant woman and a BMI model to classify as underweight ($<18.5 kg/m^2$), normal ($18.5–24.9kg/m^2$), overweight ($25-29.9kg/m^2$), obese ($30-34.9kg/m^2$) and morbidly obese($>35kg/m^2$)~\cite{kenny2013advanced}.

A report~\cite{program2000report} and a study~\cite{mustafa2012comprehensive} classify hypertensive disorders of pregnancy into following categories: gestational hypertension, chronic hypertension, pre-pre-eclampsia, and pre-eclampsia superimposed on pre-existing hypertension. According to the international guidelines, the treatment of hypertension in pregnancy varies with respect to thresholds.

Authors in the paper~\cite{douglas2011oxygen} develop a method to fix up SpO2 value for pregnant woman with pre-eclampsia by PIERS (Pre-eclampsia Integrated Estimate of RiSk). They discover threshold values of SpO2 as $\leq 93\%$ which confers the particular risk.

Diabetes in pregnancy is well studied in the paper~\cite{ali2011diabetes}, which proposes new diagnostic criteria for gestational diabetes and finds fasting plasma glucose:5.1 mmol/l, 1 h plasma glucose:10.0 mmol/l, 2 h plasma glucose:8.5 mmol/l. The authors also discuss different management procedure of diabetes and mention that diabetes may not be harmful to a pregnant woman if it can be monitor properly, otherwise, gestational diabetes may become a big issue comparatively with other risk factors of a pregnant woman.

Hyperthermia has been studied in~\cite{chambers2006risks} to demonstrate an association between high maternal fever in early pregnancy and NTDs. The author makes experiments with hot tub or spa used by pregnant women and find a threshold body temperature value and maintained below 38.9°C.

In the paper~\cite{sui2014physical}, authors survey through physical activity questionnaires and find an interesting correlation of BMI and GWE. Women with higher BMI have a larger decline in physical activity and less GWE.

%
%In the web site [The Normal Pulse Rate During Pregnancy{Livestrem.com } and {www.heartdiseaseandpregnancy.com/pat_ccdp.html }] During pregnancy, the amount of blood pumped by the heart (cardiac output) increases by 30 to 50\%. As cardiac output increases, the heart rate at rest speeds up from a normal prepregnancy rate of about 70 beats per minute to 80 or 90 beats per minute.

%In the web site [Boots WebMD Medical Reference] given lots of information about different gestational risk factor. They have research groups and linked with related resources. In this web site explore different status and symptom of First Trimester (nausea) like Lethargy, confusion, or a decreased alertness,Severe abdominal pain,Diarrhea Rapid breathing as a threshold status on the other hand also mention more symptom like blood in the vomit (bright red or "coffee grounds" in appearance) Severe headache or stiff neck. In the First Trimester (vaginal discharge) like Bloody or brown,Cloudy or yellow,Frothy, yellow or greenish with a bad smell,Pink,Thick, white, cheesy,White, gray, or yellow with fishy odor. In this web site also discuss about Third Trimester(Contraction) factors like periodic,irregular,regular.

A researcher develops Maternal Serum Screening (MSS) methods by using different case study. He has suggested MSS level for Distribution of MSAFP levels with fetal neural tube defects as 7.0 (unaffected), 2.3(Spina Bifida), 5.0(Anencephaly) and MSS level for distribution of MSAFP levels with fetal Down syndrome as 0.5/LR-2.0(Down syndrome), 0.8/LR-1.0,1.4/LR 0.5(unaffected) during the second trimester of pregnancy period~\cite{chodirker2001maternal}.

%In the paper [Abnormal Position and Presentation of the Fetus {Julie S. Moldenhauer, MD} (MSD Manual)] authors discussed about different position and presentation of fetus which is refer to risk status of pregnant women.They have suggested the following fetus position and presentation .Facing backward and Head first is normal position,Facing forward Face and brow is moderate position and position in breech, shoulder is abnormal.

In the paper~\cite{lucovnik2011use}, authors develop a new method to detect the uterine contraction by using changes of several EMG parameters. They have considered the parameters power spectrum (PS) peak frequency and propagation velocity(PV) to find out a summation value as a cut-off. They observe an interesting finding that if the value exceeds the cut-off as 84.48, the delivery will be within 7 days. They have experimented against 100 women and theoretically got a true labor time.

Fetal heart rate is observed in~\cite{von2013normal} with a methodology based on the ``delayed moving windows'' algorithm. The researchers have determined the normal fetal heart rate as 120 to 160 bpm.

%In the web site [Fetal heart rate {Dr Jeremy Jones and Dr Yuranga Weerakkody et al.} ] also founded that normal 120 to 160 beats per minute (bpm).
%In the web site [http://www.whattoexpect.com/pregnancy/fetal-movement ] describe Fetal movement activity in details. From month 4 of pregnancy to month 9 of pregnancy need to look after the fetal movement.10 movements such as kicks, flutters, or rolls. within 2 hours may be consider as threshold value.

\subsection{Synthesis Matrices}
Fig.~\ref{fig:study-selection} summarizes our process of literature search and
study selection based inclusion and exclusion criteria to select 9 prospective studies for detailed
synthesis matrices. We have selected the 9 contributions as they define a full system to address maternal 
health care issues. The synthesis matrices are developed based on the meta-information of the articles,
different systems used in their proposal, technology usage, different maternal factors and qualitative 
attributes.

We summarize the meta information of the contributed articles in Table~\ref{tab:sm:meta}, which depicts
the fact that the Cyber-Physical-System based maternal health care systems have been widely studied in 
USA, European countries and China. 

\begin {table}[htb]
\caption {Meta-information} \label{tab:sm:meta} 
\begin{center}
\begin{tabular}{|l|l|l|l|l|l|} \hline
\backslashbox{source}{attribute} & Year & Location\\\hline
mMonitoring~\cite{lyu2013multi} 	& 2013 & China\\\hline
homeCPS~\cite{horoba2014design} 	& 2014 & Poland \\\hline
wSensors~\cite{penders2015wearable} & 2015 & USA, The Netherlands, Belgium\\\hline
fMDU~\cite{wrobel2014automated} & 2014 & Poland \\\hline
mMamee~\cite{karagiannaki2015mmamee} & 2015 & Greece, UK \\\hline
bsAcquisition~\cite{pawlak2015telemonitoring} & 2015 & Poland \\\hline
health-CPS~\cite{zhang2015health} & 2015 & Saudi Arabia, China, USA, Taiwan \\\hline
hTelecare~\cite{wrobel2015medical} & 2015 & Poland \\\hline
iWSN\cite{nictulescu2015integrated} & 2015 & Romania \\
\hline
\end{tabular}
\end{center}
\end{table}

In Table~\ref{tab:sm:systems}, we have observed different studies with maternal and fetal monitoring, telemonitoring, used storage and Decision Making System (DMS) attributes to elaborate properties of each of the articles. The table depicts that all of the reviewed systems have maternal telemonitoring systems with Maternal Heart Rate (MHR), Electrocardiogram (ECG) and other common monitoring factors. However, two articles do not take fetal monitoring into account. Fetal monitoring often observes Fetal Heart Rate (FHR), ECG, Cardiotocography (CTG), QRS (three graphical deflections on ECG), fetal movement and so on. Moreover, each study focuses on different monitoring factors along with different DMS. Interestingly, almost half of the studies use cloud storage, whereas the rest half of the studies use internet enabled hospital system.

\begin {table*}[htb]
\caption {Systems} \label{tab:sm:systems} 
\begin{center}
\begin{tabular}{|l|l|l|l|l|l|} \hline
\backslashbox{source}{attribute} & \shortstack{Maternal\\ Monitoring} & \shortstack{Fetal\\ Monitoring} & \shortstack{Telemonitoring} & Storage & DMS\\\hline
mMonitoring~\cite{lyu2013multi} & continuous & ECG, CTG, STAN & yes & cloud & yes\\\hline
homeCPS~\cite{horoba2014design} & on demand 
& \shortstack{FHR, QRS, \\movement, ECG}  & yes & \shortstack{hospital \\surveillance center} & yes\\\hline
wSensors~\cite{penders2015wearable} & yes & × & yes & × 
& \shortstack{machine learning and\\ pattern recognition}\\\hline
fMDU~\cite{wrobel2014automated} & cardiotocograph & FHR, amniotic fluid & × & × & Butterworth filter\\\hline
mMamee~\cite{karagiannaki2015mmamee} & yes & × & yes & cloud & analysis and correlation\\\hline
bsAcquisition~\cite{pawlak2015telemonitoring} & GSVD 
& \shortstack{FHR, movement,\\CTG,QRS} & yes 
& \shortstack{hospital\\surveillance center} 
& \shortstack{ICA, abdominal\\signal analysis}\\\hline
health-CPS~\cite{zhang2015health} & MHR and electrocardiogram & × 
& \shortstack{individual\\vital sign} & cloud & yes\\\hline
hTelecare~\cite{wrobel2015medical} 
& \shortstack{uterine activity, blood glucose,\\ MHR, blood pressure and oxygenation}
& FHR, ECG & yes 
& \shortstack{hospital\\surveillance center} 
& \shortstack{automated\\ quantitative analysis}\\\hline
iWSN\cite{nictulescu2015integrated} 
& \shortstack{uterine contractions,respiration rate,\\Pulse oximetry, hemoglobin in blood} 
& FHR & yes & cloud & wavelet signal analysis\\
\hline
\end{tabular}
\end{center}
\end{table*}

We have attributed technology with CPS, sensors and communication technologies. Our observation on the studies is articulated in Table~\ref{tab:sm:technology}. Almost half of the studies have already started using Medical Cyber-Physical-Systems, while all of them are using different sort of sensors such as Body Area Network (BAN), Physical Area Network (PAN), Micro-Detectors (MD), Maternal ECG (MECG), Fetal ECG (FECG), electrohysterography (EHG) and so on. Interestingly, bluetooth is quite common as a sensor communication technology. 

\begin {table*}[htb]
\caption {Technology} \label{tab:sm:technology} 
\begin{center}
\begin{tabular}{|l|l|l|l|} \hline
\backslashbox{source}{attribute} & \shortstack{CPS} & \shortstack{Sensors} & Communication\\\hline
mMonitoring~\cite{lyu2013multi} 		& × 	& \shortstack{Wearable Wireless} & \shortstack{serial port, bluetooth, Zigbee,\\ RF433, smart-phone} \\\hline
homeCPS~\cite{horoba2014design} 		& MCPS 	& \shortstack{BAN, PAN, maternal and fetal,\\external and environmental} & \shortstack{bluetooth, Zigbee, GSM,\\ Internet PDA, Laptop} \\\hline
wSensors~\cite{penders2015wearable} 	& × 	& \shortstack{Mobile and wearable sensors} & ×\\\hline
fMDU~\cite{wrobel2014automated} 		& × 	& \shortstack{Doppler ultrasound } & × \\\hline
mMamee~\cite{karagiannaki2015mmamee} 	& × 	& \shortstack{Smart sensors} & ZigBee, bluetooth \\\hline
bsAcquisition~
\cite{pawlak2015telemonitoring} 		& MCPS 	& \shortstack{BAN,PAN, biosignal networked \\MD, MECG, FECG, EHG} & ZigBee, bluetooth\\\hline
health-CPS~\cite{zhang2015health} 		& MCPS 	& \shortstack{Wearable BAN} & Internet, bluetooth\\\hline
hTelecare~\cite{wrobel2015medical} 	& MCPS 	& \shortstack{BAN, PAN, MD} & WAN/Internet\\\hline
iWSN\cite{nictulescu2015integrated} 	& × 	& \shortstack{mobile cardiotocograph and \\body sensors,  Lilypad} & \shortstack{Internet, wireless network,\\ bluetooth, smart-phone}\\
\hline
\end{tabular}
\end{center}
\end{table*}

There are a number of factors that can affect maternal health. These factors are characterized as external, environmental, risk, physical and behavioral factors. External factors often address eating habits, maternal activities, and the geographic location at which a pregnant woman is staying instantaneously while environmental factors are temperature, humidity, noise, the amount of $CO_2$ in the atmosphere, water, dust and etc. Risk factors are the main focusing points for maternal pregnancy risk calculation. However, behavioral factors such as sleeping pattern, stress, diet and weight management, smoking and drinking have hazardous relation with fetal development and maternal health. Our observation on different studies compares the facts in Table~\ref{tab:sm:factors}.

\begin {table*}[htb]
\caption {Factors} \label{tab:sm:factors} 
\begin{center}
\begin{tabular}{|l|l|l|l|l|l|} \hline
\backslashbox{source}{attribute} & \shortstack{External} & Environmental & Risks & \shortstack{Physical\\activity} & Behavioral\\\hline
mMonitoring~\cite{lyu2013multi} 	& ×	& × & \shortstack{ECG, blood glucose,\\ blood pressure, oximeter,\\
weight, and fat monitors} & × & ×\\\hline
homeCPS~\cite{horoba2014design} 	& GPS,maternal activity & temperature, humidity & 
\shortstack{blood pressure, oxygenation,\\glucose, body temperature,\\ activity, maternal ECG, EUC,\\ MUC, FHR, fetal ECG,\\ fetal movement} & yes & ×\\\hline
wSensors~\cite{penders2015wearable} & yes & × & \shortstack{obesity, BP, BG, stress,\\ anxiety, sleep disorder} & yes & 
\shortstack{sleep~\cite{pien2004sleep}, stress~\cite{huizink2003stress},\\
diet and weight management~\cite{thangaratinam2012interventions,louie2011randomized},\\ smoking~\cite{sexton1984clinical}, and drinking~\cite{carson2010alcohol}}\\\hline
fMDU~\cite{wrobel2014automated} & × & × & fetal movement activity & × & ×\\\hline
mMamee~\cite{karagiannaki2015mmamee} & GPS, environmental sensor & \shortstack{temperature, humidity,\\ $CO_2$, noise, water, dust} & external factors & × & ×\\\hline
bsAcquisition~\cite{pawlak2015telemonitoring} & GPS, maternal activity & temperature, humidity & 
\shortstack{diabetes, post-term pregnancy,\\ pregnancy induced hypertension} & × & ×\\\hline
health-CPS~\cite{zhang2015health} & eating habits & temperature, humidity & obesity and high blood pressure & yes & emotion\\\hline
hTelecare~\cite{wrobel2015medical} & GPS, maternal activity & temperature, humidity & 
\shortstack{diabetes, pregnancy\\ induced hyper tension,\\ post-term pregnancy, BMI} & yes & behavioral model\\\hline
iWSN\cite{nictulescu2015integrated} & × & × & hypoxia & yes & ×\\
\hline
\end{tabular}
\end{center}
\end{table*}

At last but not least the synthesis matrix focuses on the qualitative analysis of our reviewed articles. Among the qualitative factors, privacy stays on the top to make such personal health care system successful. As the health care is a critical issue, it needs a high level of safety, security and reliability. Moreover, as the maternal health care has a dynamic behavior over time, telemonitoring often requires self-adaptability to make this a success. After all, maternal health care often requires many sensors, technologies and systems to work together. Therefore, interoperability plays an important role in this critical issue. The comparison has been depicted in Table~\ref{tab:sm:qualitative} to point on some interesting research gaps.

\begin {table*}[htb]
\caption {Qualitative Analysis} \label{tab:sm:qualitative} 
\begin{center}
\begin{tabular}{|l|l|l|l|l|l|} \hline
\backslashbox{source}{attribute} & Privacy & Security & Interoperability & Adaptability & Reliability\\\hline
mMonitoring~\cite{lyu2013multi} 		& × & × & yes & × & yes\\\hline
homeCPS~\cite{horoba2014design} 	& yes & × & yes & × & yes\\\hline
wSensors~\cite{penders2015wearable} & yes & × & × & yes & yes\\\hline
fMDU~\cite{wrobel2014automated} & × & × & × & × & ×\\\hline
mMamee~\cite{karagiannaki2015mmamee} & yes & × & yes & yes & yes\\\hline
bsAcquisition~\cite{pawlak2015telemonitoring} & × & × & yes & yes & yes\\\hline
health-CPS~\cite{zhang2015health} & yes & security tag & × & yes & yes\\\hline
hTelecare~\cite{wrobel2015medical} & × & yes & yes & yes & yes\\\hline
iWSN\cite{nictulescu2015integrated} & × & × & × & × & ×\\
\hline
\end{tabular}
\end{center}
\end{table*}

\subsection{Other contributions}

Our study also observes the strength of a few other articles in the discourse such that the strength can be addressed during designing a new system.

Authors in the article~\cite{banerjee2012ensuring} articulate S3, {\em defined as safety, security and sustainability}, where safety focuses on the avoidance of hazards, security on the assurance of
integrity, authenticity, and confidentiality of information, while sustainability addresses the 
maintenance of a long-term operation of CPSs using green sources of energy. 

The article~\cite{wang2011secured} introduces CPeSC3 (Cyber Physical enhanced Secured wireless sensor networks in 3 Cores) using integrated cloud computing in a medical health care application scenario focusing on 3C defined as, 1.communication core, 2.computation core and 3. resource scheduling and management core.

Privacy issue in maternal health care system has a prevalent impact on societal acceptance. This issue has been widely studied in~\cite{talpur2014internet}. This study focuses on the services for pregnant women at anytime and anywhere, with anything and anyone, where the actors are pregnant women, gynecologists, and caregivers.

The article~\cite{chen2011intelligent} focuses on a {\em real-time} intelligent system that uses data mining technique against continuous sensor data kept in data center to produce service delivery. This study takes distributed file system and privacy issue into account.

In the article~\cite{gorthi2009automated}, authors address a machine learning approach for the early detection of the pregnancy risk based on patterns gleaned from a profile of known clinical parameters. It can predict and detect early complications of pregnancy using clinical decision support system (CDSS), which uses classification and regression trees to evaluate the risk category of a pregnancy.

\section{Discussion}
\label{sec:discussion}

As we have concentrated our study on maternal health care, we observe the fundamental risk factors of maternal health care. An author group, K. Horoba et al have done a number of research on the maternal health care considering Medical CPS and addressing important risk factors related to a pregnant woman. To develop an MCPS based monitoring system for maternal health care, it is always necessary to consider the system as a critical application. Every pregnant woman may have any two, more risk complexities, which are usually combined to produce a final alert to the patient as well as to the care -providers. As a critical domain, it often needs an intervention of medical professionals.

Through our analytical review over risk factors, we have found a research paper~\cite{gorthi2009automated} that focuses on the automatic risk assessment tool for pregnancy care using Clinical Decision Support System (CDSS). They have addressed a process of combining pregnancy risk factors to classify synthetically the instantaneous pregnancy states as normal, moderate or high-risk pregnancies. Our study also focuses on all of the respective risk factors to find an interesting correlation.

Patients of pregnancy are called to be at high-risk if she or her baby has an increased chance of a health problem. There are thousands of risk factors for pregnant woman categorized as physical, behavioral, environmental, lifestyle factors. As our goal of this work is to study on maternal health care targeting Cyber-Physical-Systems, we analyzed especially the risk factors related to the pregnancy. 

The recent advancement of medical devices, intelligent sensors, Internet of Things (IoT), efficient telecommunication and information based smart decision support system (DSS) has paved a paradigm shift in maternal health care. However, it necessitates the data filtering at a physical level, automatic integration of filtered information before taking a decision, avoiding data hazards, assurance of integrity, authenticity, and confidentiality of information, data failure prevention, sustainability using green technology are the key ideas to make a great success of the critical mission in maternity health care telemonitoring systems targeting Medical Cyber-Physical-Systems. 

\section{Conclusion}
\label{sec:conclusion}

In summary of this review paper, our observations were of three folds. We observed different risk factors associated to maternal health care during pregnancy, while we studied comparisons of a few proposed systems considering meta-information of the articles, system, technology, factors and qualitative attributes. Moreover, we observed a few other well-designed prospective studies with the longitudinal assessment to find out some more interesting attributes to be focused on Medical Cyber-Physical-Systems (CPS) based maternal health care telemonitoring system.

%link up of abstract in conclusion

%what you have focused

%what you have achieved

%what is the merit

%future target

\section*{Acknowledgement}
This work was supported by the European Erasmus Mundus cLINK programme at the University Lumiere Lyon 2, France.

\bibliographystyle{IEEEtran}

\bibliography{mhc}

\end{document}